\documentclass[aps,showpacs,pra,preprint]{revtex4}
\usepackage{mathrsfs} 
\usepackage{amssymb}

\usepackage{graphicx}
\usepackage{dcolumn}
\usepackage{bm}

\begin{document}

\title{Non-Markovian Dynamics of Quantum and Classical Correlations in the Presence of System-Bath Coherence}
\author{ Chuan-Feng Li$\footnote{
email: cfli@ustc.edu.cn}$, Hao-Tian Wang, Hong-Yuan Yuan, Rong-Chun
Ge, and Guang-Can Guo} \affiliation{Key Laboratory of Quantum
Information, University of Science and Technology of China, CAS,
Hefei, 230026, People's Republic of China}
\date{\today }


\begin{abstract}
We present a detailed study of the dynamics of correlations in
non-Markovian environments, applying the hierarchy equations
approach. This theoretical treatment is able to take the system-bath
interaction into consideration carefully. It is shown that crosses
and sudden changes of classical and quantum correlations can happen
if we gradually reduce the strength of the interactions between
qubits. For some special initial states, sudden transitions between
classical and quantum correlations even occur.
\end{abstract}

\maketitle

\section{introduction}
In the field of quantum information and computation, a great amount
of meaningful investigations have been done to explore the dynamics
of two-level systems in environments with different properties
\cite{1,2,3,4}. Information extracted from a bipartite system comes
to the center of our interests because in this simplest scenario we
can not only conveniently introduce our theoretical calculations but
also obtain the same results even with different definitions. In a
two-qubit system, quantum mutual information has been defined as the
largest amount of information which we can extract between the
qubits \cite{5.7}. This mutual information consists of two different
parts, the classical correlation and the quantum correlation.
Quantum correlation, which is equivalent to quantum discord in some
special cases \cite{1.2}, can also be divided as quantum
entanglement and non-entanglement quantum correlation \cite{6.29}.
Many works have been focused on the different dynamics between
quantum correlation and entanglement \cite{1}.

The most popular definitions of quantum information are discussed by
Vedral, in which the total correlations (quantum mutual information)
comes to the form as \cite{6.28}
\begin{eqnarray}
\mathcal{I}(\rho_{AB})=S(\rho_A)+S(\rho_B)-S(\rho_{AB}),
\end{eqnarray}
where $\rho_{A(B)}$ is the reduced density matrix of $\rho_{AB}$ and
$S(\rho)$ represents the von Neumann entropy which can be calculated
as $S(\rho)=-\Sigma_i\lambda_i\log_2\lambda_i$, in which $\lambda_i$
is the eigenvalue of $\rho$. Classical correlation can be obtained
from the measurements of subsystem $B$ under the complete set of
orthonormal projectors in the form
\begin{eqnarray}
\Pi_\parallel=\sin^2\theta|1\rangle\langle1|+\cos^2\theta|0\rangle\langle0|+\sin\theta\cos\theta(e^{i\phi}|1\rangle\langle0|+e^{-i\phi}|0\rangle\langle1|)
\end{eqnarray}
and
\begin{eqnarray}
\Pi_\perp=\cos^2\theta|1\rangle\langle1|+\sin^2\theta|0\rangle\langle0|-\sin\theta\cos\theta(e^{i\phi}|1\rangle\langle0|+e^{-i\phi}|0\rangle\langle1|),
\end{eqnarray}
where $|1\rangle$ and $|0\rangle$ are the excited and ground states
of the qubit. Then classical correlation can be expressed as
\cite{2,6.6}
\begin{eqnarray}
\mathcal{C}(\rho_{AB})=\max_{\theta,\phi}[S(\rho_A)-\Sigma_jq_jS(\rho_A^j)]
\label{6}
\end{eqnarray}
with
$\rho_A^j=\mathrm{Tr}_B[(\mathbb{I}\otimes\Pi_j)\rho_{AB}(\mathbb{I}\otimes\Pi_j)]/q_j$,
$q_j=\mathrm{Tr}_{AB}[\rho_{AB}(\mathbb{I}\otimes\Pi_j)]$, and
$j=\parallel, \perp$. Since the total correlations can also be
expressed as
\begin{eqnarray}
\mathcal{I}(\rho_{AB})=\mathcal{C}(\rho_{AB})+\mathcal{Q}(\rho_{AB})
\end{eqnarray}
then the quantum correlation is
\begin{eqnarray}
\mathcal{Q}(\rho_{AB})=\mathcal{I}(\rho_{AB})-\mathcal{C}(\rho_{AB}).
\end{eqnarray}

With above basic equations, the left work is to give the values of
the system matrix $\rho_{AB}$. In most scenarios we may find that
the classical and quantum correlation would have mediocre behaviors,
such as the result that classical correlation is always lager than
quantum correlation. However, Many meaningful theoretical and
experimental works have been done recently to demonstrate that with
some special channels and a class of initial states, the classical
correlation and quantum correlation will both remain constant in
different regimes, and there is a sudden transition from the
classical to the quantum decoherence \cite{5,6,7}.

In this work we demonstrate that in some common cases we can also
obtain crosses of classical and quantum correlations and correlation
sudden changes. By applying the hierarchy equations approach
\cite{4,10} we set up the model of a two interacting qubits system
in independent non-Markovian environments, and we also carefully
take into account the presence of system-bath coherence, which has
been proved to dramatically influence the dynamics of the quantum
system. As the strength of the interaction between the two qubits
decreases, we can find the crosses and sudden changes gradually
appear. When there are not interactions at all, the changes will be
the sharpest.

\section{The Model and Theory}
We present a newly developed theoretical method called hierarchy
equations approach to handle the system dynamics \cite{4,10}. This
method can conveniently treat the scenario in which there is a
non-Markovian environment. We know that both the quantum system and
the heat bath have characteristic frequencies. Most of theoretical
treatments would suppose a fast bath to make the Born and ultrafast
bath approximations valid \cite{4}. However, in this hierarchy
equations approach, a slow noise bath can be applied. This kind of
environment will record part of the information from the system to
the bath and then return some to the system.

We set $\hbar=1$ throughout this report. The environment is divided
into two independent baths interacting with two interacting qubits.
Then we write the standard system Hamilton as
\begin{eqnarray}
H_S=\varepsilon(a_1^{\dag}a_1+a_2^{\dag}a_2)+\zeta(a_1^{\dag}+a_1)(a_2^{\dag}+a_2),
\end{eqnarray}
where $\varepsilon$ is the energy gap of the two-level system. The
subscripts $1$ and $2$ represent the two qubits, $a^{\dag}$ and $a$
are creation and annihilation operators. $\zeta$ stands for the
strength of the interactions between two qubits, so we can
conveniently reduce its value until zero to observe the changes of
the correlations. The heat bath consists of a set of oscillators,
with the form \cite{3,4,3.26}
\begin{eqnarray}
H_{B+SB}=\sum_j(\frac{p_j^2}{2m_j}+\frac{1}{2}m_j\omega_j^2x_j^2)-\sum_{m,j}C_{mj}(a_m^{\dag}+a_m)x_j,
\end{eqnarray}
$x_j$, $p_j$, $m_j$ and $\omega_j$ are coordinate, momentum, mass
and frequency of the $j$th harmonic oscillator, respectively. Then
$H_B$ presents the Hamilton of the bath and $H_{SB}$ is the
interaction between system and environment. $C_{mj}$ stands for the
coupling strength of the $j$th oscillator to the $m$th qubit. All of
the system-bath coupling information can be found though the
spectral density function $J(\omega)$ and in this work we model it
as the Lorentzian cutoff
\begin{eqnarray}
J(\omega)=\frac{\omega\eta\gamma}{\omega^2+\gamma^2}.
\end{eqnarray}
$\eta$ represents the system-bath coupling strength, and $\gamma$ is
the characteristic frequency of the heat bath. When the fast bath
condition is fulfilled as $\hbar\gamma\gg\varepsilon$, we can choose
the initial state in which the system and bath are independent. If
$\hbar\gamma\sim\varepsilon$ it represents a non-Markovian
environment and that is just our scenario. In the present work we
choose $\varepsilon=1.5\delta$ and $\gamma=4\delta$, and here
$\delta$ stands for an energy unit. With a Fourier transform of the
spectral density function we have
\begin{eqnarray}
F(t)=\sum_{k=0}^{\infty}c_ke^{-\gamma_k|t|},
\label{10}
\end{eqnarray}
where $\gamma_k$ is a Bosonic Matsubara frequency, with
$\gamma_0=\gamma$ and $\gamma_k=2{\pi}k/\beta$ when $k\geq1$. The
coefficients in front of the exponential function in Eq.\,\ref{10}
have values as $c_0=\frac{\eta\gamma}{2}(-i+\cot\beta\gamma/2)$ and
$c_k=2\eta\gamma_0\gamma_k/\beta(\gamma_k^2-\gamma_0^2)$ when
$k\geq1$. The equation about the dynamics of the system using
hierarchy equations approach, witch has the following form \cite{4}

\begin{eqnarray}
\nonumber\\
\frac{d\rho_{\mathbf{n}}(t)}{dt}&=&-(i\mathcal{L}+\sum_{m=1}^2\sum_{k=0}^Kn_{mk}\gamma_k)\rho_{\mathbf{n}}(t)\nonumber-\sum_{m=1}^2((\frac{1}{\beta\gamma_0}-i\frac{1}{2})\eta-\sum_{k=0}^K\frac{c_k}{\gamma_k})[V_m,[V_m,\rho_{\mathbf{n}}(t)]]\nonumber\\
&&-i\sum_{m=1}^2\sum_{k=0}^Kn_{mk}(c_kV_m\rho_{n_{mk}-1}(t)-c_k^\ast\rho_{n_{mk}-1}(t)V_m)\nonumber\\
&&-i\sum_{m=1}^2\sum_{k=0}^K[V_m,\rho_{n_{mk}+1}(t)],
\label{11}
\end{eqnarray}
where $\mathcal{L}$ is the Liouvillian operator of the two-level
system with $\mathcal{L}\rho=[H_S,\rho]$, and $V_m=a_m^{\dag}+a_m$
with $m=1, 2$.

In Eq.\,\ref{11} $\rho_\mathbf{n}$ are called auxiliary density
operators (ADOs) \cite{3}. The subscript $\mathbf{n}$ is a
multi-index which can be extended as $n_{mk}$, where $m$ represents
the number of qubits in the system and the $k$ represents that of
the Bosonic Matsubara frequencies. The notation $n_{mk}\pm1$ refers
to an increase and decrease of this index. Only when all the values
of the multi-index are zero it stands for the physical system
density operator. And if one of the notation $n_{mk}\pm1$ has a
value under zero or the sum of all the notations exceed a limit
\cite{10}, the density operator is set to be zero. The values in the
ADOs contain a deal of important information about the system-bath
interaction, which has been proved to be able to dramatically change
the dynamics of the system. With a carefully consideration of the
system-bath coupling we find that it is a important factor which can
greatly influence our results of the system evolution. Another
advantage of this theoretical treatment is the convenience of
applying the method into computer programs. It is quite easy to
numerically calculate the dynamics of the system density matrix
though a program. This hierarchy equations approach will be a useful
treatment in scenarios when the interaction between the system and
the environment is very strong.

\section{The Result and Discussion}
In this work the environment is non-Markovian and much more common
than the ones in Ref.\,\cite{1,7}. We set $\eta=0.3\delta$,
$\gamma=4\delta$ and $\beta=2.5/\delta$. The initial state of the
two-qubit system is a Bell state with
$|\psi\rangle=1/\sqrt{2}(|10\rangle-|01\rangle)$. As the system
Hamilton only has diagonal and anti-diagonal elements, the system
density matrix will evolve into $X$ states \cite{8}, which can
represent a great number of quantum systems. We use numerical method
to calculate the mutual information, classical correlation and
quantum correlation with enough precision.

\begin{figure}[tbph]
\centering
\includegraphics[width= 3in]{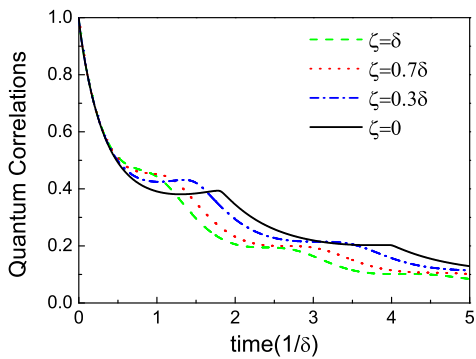}
\caption{(color online) The green dashed, red dotted, blue
dash-dotted and black solid lines represent the quantum correlations
as functions of time with $\zeta=\delta$, $0.7\delta$, $0.3\delta$
and zero, respectively.}\label{1}
\end{figure}

\begin{figure}[tbph]
\centering
\includegraphics[width= 3in]{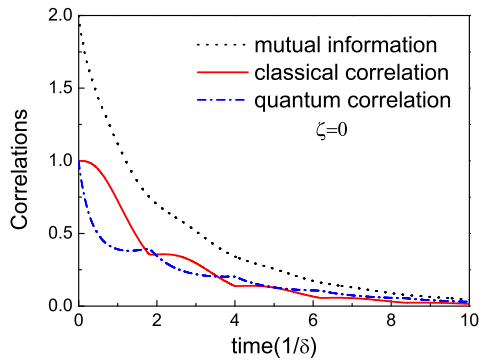}
\caption{(color online) The black dotted, red solid and blue
dash-dotted lines represent the mutual information, classical
correlation and quantum correlation as functions of time,
respectively.}\label{2}
\end{figure}
In order to show the correlation sudden changes, we set
$\zeta=\delta, 0.7\delta, 0.3\delta$ and $0$ respectively in
Figure\,\ref{1}. Obviously, with the decrease of $\zeta$, the
quantum correlation becomes sharper and sharper. When $\zeta=0$, a
correlation sudden change occurs at certain time. For a complete
view of this interesting change, we show the whole evolution of
mutual information, classical correlation and quantum correlation
with no interactions between the qubits in Fig.\,\ref{2}. The mutual
information goes down monotonically as the system evolves. However,
it is shown clearly that there are several crosses and sudden
changes of the classical and quantum correlation. At first the
quantum correlation shows a high speed of decay as a result of the
interaction with the bath. Then the decay stops and the quantum
correlation even shows a little increase, which clearly demonstrate
the non-Markovian memory effects. However, the classical decoherence
speeds up and at around $t=1.7/\delta$ these two correlations reach
the same value. A cross immediately after the time shows the lager
discord than the classical correlation during a short interval.
Another cross emerges at $t\approx3.6/\delta$. After this we may not
find any crosses and the quantum correlation is then always larger
than classical correlation. With careful observation we know that at
both crosses and some other time points (such as
$t\approx6.2/\delta$ and $t\approx8.4/\delta$) there are sudden
changes of the correlations. These sudden changes catch interests
because they may be results of some quantum phase transitions or the
transition of classical and quantum decoherence \cite{7}. We also
find that the length of the intervals between these changes are all
nearly $2.2/\delta$. This phenomenon may indicate some mechanisms of
the system evolution in a common non-Markovian environment.

\begin{figure}[tbph]
\centering
\includegraphics[width= 3in]{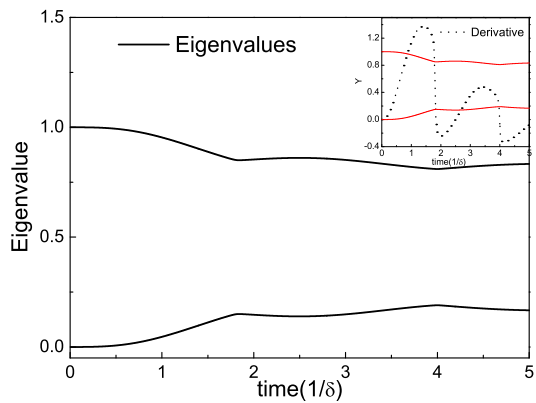}
\caption{(color online) The eigenvalues of the density operator
$\rho_A^j$ after the maximization procedure as a function of time.
The dotted line in the inserted figure at the top right corner is
the derivative of the lower eigenvalue.}\label{3}
\end{figure}

In order to investigate the sudden changes, we calculate the
eigenvalues of $\rho_A^j$ after the maximization procedure in
Eq.\,\ref{6}. Fig.\,\ref{3} shows the first two sudden changes of
the eigenvalues. These two curves have exactly the opposite trends
and both have two inflectional points, at which the derivative can
become quite sharp, as we show in the inserted figure. Obviously the
two points correspond to the sudden changes in the classical and
quantum correlations.

\begin{figure}[tbph]
\centering
\includegraphics[width= 5.5in]{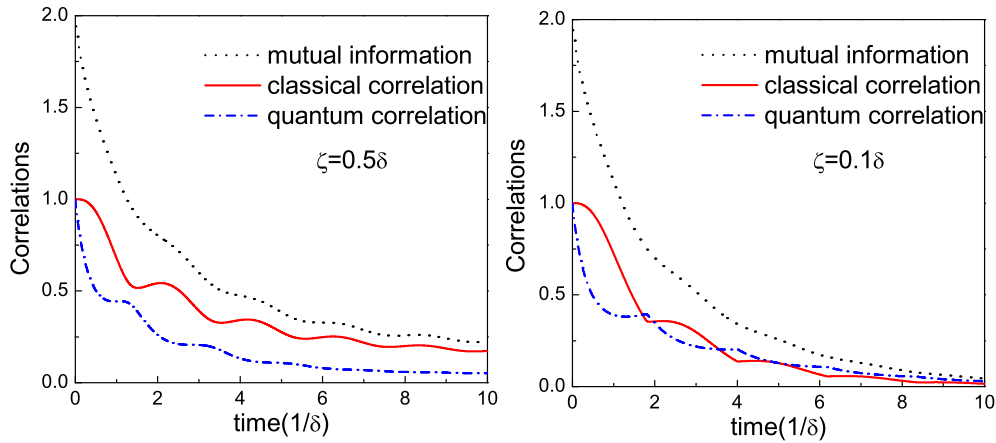}
\caption{(color online) Mutual information, classical correlation
and quantum correlation for different coupling strength between the
qubits. $\zeta=0.5\delta$ in the left figure and $\zeta=0.1\delta$
in the right one.}\label{4}
\end{figure}

Next we concentrate on the influence of coupling strength. We choose
$\zeta=0.5\delta$ and $\zeta=0.1\delta$ for comparison which are
shown in Fig.\,\ref{4}. The left figure shows a quite common
dynamics of the correlations, with classical correlation always
larger than quantum counterpart. When the strength of the
interaction between the two qubits decreases to $0.1\delta$, crosses
begin to emerge as the quantum correlation rises and classical
correlation goes down. However, with careful observation we find
that the changes are not so sharp as the ones in Fig.\,\ref{1} where
$\zeta=0$. So we can infer that at the point of $\zeta=0$ some
interesting phenomenon may happen such as the correlation sudden
changes.
\begin{figure}[t]
\centering
\includegraphics[width= 3in]{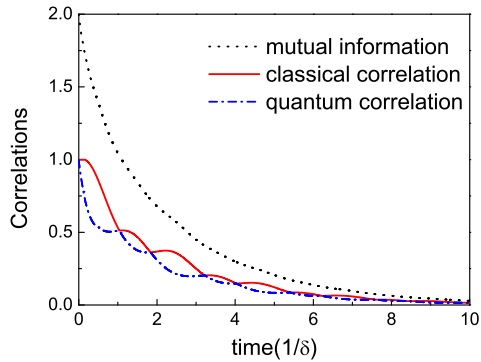}
\caption{(color online) The black dotted, red solid and blue
dash-dotted lines represent the mutual information, classical
correlation and quantum correlation as functions of time,
respectively, with an initial state of
$|\psi\rangle=1/\sqrt{2}(|11\rangle-|00\rangle)$.}\label{5}
\end{figure}

Apart from the effects of the interactions between the two qubits,
the initial states of the system also play an important role on the
evolution of the correlations, as shown in Fig.\,\ref{5}. We change
the initial state from odd
($|\psi\rangle=1/\sqrt{2}(|10\rangle-|01\rangle)$) to even parity
($|\psi\rangle=1/\sqrt{2}(|11\rangle-|00\rangle)$). The parameters
are chosen to be the same as those in Fig.\,\ref{2}. Sudden
transitions between quantum and classical correlations can be found
which has been demonstrated in other systems \cite{6,7}. This kind
of transition can occur several times in our system during the
evolution such as at the time $t=1/\delta$, $t=2/\delta$ and so on.

\section{conclusion}
We study the evolution of the classical and quantum correlations of
a two-qubit system in independent non-Markovian environments,
applying the newly developed hierarchy equations approach. The
influence of initial state and coupling strength between the qubits
are investigated. Crosses and sudden changes of classical and
quantum correlations are illustrated. We further show sudden
transition between classical and quantum correlations for certain
initial state.

\section{acknowledgement}
This work was supported by the National Fundamental Research Program
and National Natural Science Foundation of China (Grant Nos.
60921091, 10874162 and 10734060).

\end{document}